\documentstyle[11pt, leqno]{article}
\input amssym.def
\input amssym.tex
\def\unit{\one}

\def\utw{\smash{\rlap{\lower5pt\hbox{$\sim$}}}}
\def\udtw{\smash{\rlap{\lower6pt\hbox{$\approx$}}}}

\def\bbbr{{\rm I\!R}} 
\def\bbbn{{\rm I\!N}} 

\def\bbbone{{\mathchoice {\rm 1\mskip-4mu l} {\rm 1\mskip-4mu l}
{\rm 1\mskip-4.5mu l} {\rm 1\mskip-5mu l}}}
\def\bbbc{{\mathchoice {\setbox0=\hbox{$\displaystyle\rm C$}\hbox{\hbox
to0pt{\kern0.4\wd0\vrule height0.9\ht0\hss}\box0}}
{\setbox0=\hbox{$\textstyle\rm C$}\hbox{\hbox
to0pt{\kern0.4\wd0\vrule height0.9\ht0\hss}\box0}}
{\setbox0=\hbox{$\scriptstyle\rm C$}\hbox{\hbox
to0pt{\kern0.4\wd0\vrule height0.9\ht0\hss}\box0}}
{\setbox0=\hbox{$\scriptscriptstyle\rm C$}\hbox{\hbox
to0pt{\kern0.4\wd0\vrule height0.9\ht0\hss}\box0}}}}
\def\bbbe{{\mathchoice {\setbox0=\hbox{\smalletextfont e}\hbox{\raise
0.1\ht0\hbox to0pt{\kern0.4\wd0\vrule width0.3pt
height0.7\ht0\hss}\box0}}
{\setbox0=\hbox{\smalletextfont e}\hbox{\raise
0.1\ht0\hbox to0pt{\kern0.4\wd0\vrule width0.3pt
height0.7\ht0\hss}\box0}}
{\setbox0=\hbox{\smallescriptfont e}\hbox{\raise
0.1\ht0\hbox to0pt{\kern0.5\wd0\vrule width0.2pt
height0.7\ht0\hss}\box0}}
{\setbox0=\hbox{\smallescriptscriptfont e}\hbox{\raise
0.1\ht0\hbox to0pt{\kern0.4\wd0\vrule width0.2pt
height0.7\ht0\hss}\box0}}}}
\def\bbbq{{\mathchoice {\setbox0=\hbox{$\displaystyle\rm Q$}\hbox{\raise
0.15\ht0\hbox to0pt{\kern0.4\wd0\vrule height0.8\ht0\hss}\box0}}
{\setbox0=\hbox{$\textstyle\rm Q$}\hbox{\raise
0.15\ht0\hbox to0pt{\kern0.4\wd0\vrule height0.8\ht0\hss}\box0}}
{\setbox0=\hbox{$\scriptstyle\rm Q$}\hbox{\raise
0.15\ht0\hbox to0pt{\kern0.4\wd0\vrule height0.7\ht0\hss}\box0}}
{\setbox0=\hbox{$\scriptscriptstyle\rm Q$}\hbox{\raise
0.15\ht0\hbox to0pt{\kern0.4\wd0\vrule height0.7\ht0\hss}\box0}}}}
\def\bbbt{{\mathchoice {\setbox0=\hbox{$\displaystyle\rm
T$}\hbox{\hbox to0pt{\kern0.3\wd0\vrule height0.9\ht0\hss}\box0}}
{\setbox0=\hbox{$\textstyle\rm T$}\hbox{\hbox
to0pt{\kern0.3\wd0\vrule height0.9\ht0\hss}\box0}}
{\setbox0=\hbox{$\scriptstyle\rm T$}\hbox{\hbox
to0pt{\kern0.3\wd0\vrule height0.9\ht0\hss}\box0}}
{\setbox0=\hbox{$\scriptscriptstyle\rm T$}\hbox{\hbox
to0pt{\kern0.3\wd0\vrule height0.9\ht0\hss}\box0}}}}
\def\bbbs{{\mathchoice
{\setbox0=\hbox{$\displaystyle     \rm S$}\hbox{\raise0.5\ht0\hbox
to0pt{\kern0.35\wd0\vrule height0.45\ht0\hss}\hbox
to0pt{\kern0.55\wd0\vrule height0.5\ht0\hss}\box0}}
{\setbox0=\hbox{$\textstyle        \rm S$}\hbox{\raise0.5\ht0\hbox
to0pt{\kern0.35\wd0\vrule height0.45\ht0\hss}\hbox
to0pt{\kern0.55\wd0\vrule height0.5\ht0\hss}\box0}}
{\setbox0=\hbox{$\scriptstyle      \rm S$}\hbox{\raise0.5\ht0\hbox
to0pt{\kern0.35\wd0\vrule height0.45\ht0\hss}\raise0.05\ht0\hbox
to0pt{\kern0.5\wd0\vrule height0.45\ht0\hss}\box0}}
{\setbox0=\hbox{$\scriptscriptstyle\rm S$}\hbox{\raise0.5\ht0\hbox
to0pt{\kern0.4\wd0\vrule height0.45\ht0\hss}\raise0.05\ht0\hbox
to0pt{\kern0.55\wd0\vrule height0.45\ht0\hss}\box0}}}}

\def\bbbz{{\mathchoice {\hbox{$\sf\textstyle Z\kern-0.4em Z$}}
{\hbox{$\sf\textstyle Z\kern-0.4em Z$}}
{\hbox{$\sf\scriptstyle Z\kern-0.3em Z$}}
{\hbox{$\sf\scriptscriptstyle Z\kern-0.2em Z$}}}}

\def\diameter{{\ifmmode\oslash\else$\oslash$\fi}}

\def\init{\setcounter{equation}{0}}

\newtheorem{theoreme}{Theorem }[section]
\newtheorem{proposition}[theoreme]{Proposition}
\newtheorem{lemma}[theoreme]{Lemma}
\newtheorem{definition}[theoreme]{Definition}

\def\rr{\bbbr}

\def\nn{\bbbn}
\def\zz{\bbbz}
\def\one{\bbbone}

\def\e{{\rm e}}

\def\d{{\rm d}}
\def\12{\frac{1}{2}}

\def\proof{{\bf  Proof. }}

\def\cH{{\cal H}}
\def\ln{{\rm ln \, }}

\def\cF{{\cal F}}

\def\e{{\rm e}}

\def\pfi2{P(\varphi)_{2}}

\newcommand{\beq}{\begin{equation}}
\newcommand{\eeq}{\end{equation}}
\newcommand{\bet}{\begin{theoreme}}
\newcommand{\eet}{\end{theoreme}}
\newcommand{\bel}{\begin{lemma}}
\newcommand{\eel}{\end{lemma}}
\newcommand{\bep}{\begin{proposition}}
\newcommand{\eep}{\end{proposition}}
\newcommand{\ear}{\end{array}}

\def\cA{{\cal A}}

\begin{document}

\title{Stability and Related Properties of 
\goodbreak
Vacua and Ground States   
 \protect\footnotetext{AMS 1991 {\it{Subject 
Classification}}. 81T08, 82B21, 82B31, 46L55} \protect}          

\author{Walter F.~Wreszinski\footnote{wreszins@fma.if.usp.br} \\
Departamento de Fisica Matem\'atica \\ 
Universidade de S\~ao Paulo, Brasil \\
\and Christian D. J\"akel\footnote{christian.jaekel@mac.com} \\
Instituto de Matem\'atica y Fisica \\ Universidad de Talca, Chile
}        
\date{September 18, 2006}
\maketitle
\abstract{We consider the formal non relativistc limit (nrl) of the $:\phi^4:_{s+1}$ relativistic quantum field theory (rqft), where $s$ is the space dimension. Following work of R.~Jackiw, we show that, for $s=2$ and a given value of the ultraviolet cutoff $\kappa$, there are two ways to perform the nrl: i.)~fixing the renormalized mass $m^2$ equal to the bare mass $m_0^2$;
ii.)~keeping the renormalized mass fixed and different from the bare mass $m_0^2$.
In the (infinite-volume) two-particle sector the scattering amplitude tends to zero as $\kappa \to \infty$ in case i.) and, 
in case ii.), there is a bound state, indicating that the interaction potential is attractive. 
As a consequence, stability of matter fails for our boson system. 
We discuss why both alternatives do not reproduce the low-energy behaviour of the full rqft. The singular nature of the nrl is also nicely illustrated for $s=1$ by a rigorous stability/instability result of a different nature.}


\section{Introduction and Summary}
\init\label{introd}
\noindent
The most fundamental approach to nonrelativistic quantum mechanics is to view it as an effective theory which emerges from relativistic quantum field theory in case the initial conditions place the experimental setup in the low energy region, where both theories should give identical predictions concerning the expectation values of observable quantities like local energy densities or scattering cross-sections. Given a relativistic quantum field theory (rqft),
the non-relativistic limit (nrl), i.e., the formal limit $c\to \infty$, is used to identify the corresponding
non-relativistic theory. In spite of the conceptual importance of this topic, very few papers have been devoted to the rigorous study
of the non-relativistc limit: the classic paper of Hunziker on the nrl of the Dirac equation \cite{H}, and 
Dimock's proof of the nrl of the $:\phi^4:_2$-theory in the two-particle sector \cite{D}.

In this paper we address the problem of thermodynamic stability in the formal nrl of some 
relativistic quantum field theories, as well as other related issues concerning qualitative differences between
relativistic vacua and the ground states of the non-relativistic many particle systems. 

In Section 2 necessary notation and definitions
pertaining to the $:\phi^4:_{s+1}$ theories and their formal nrl are introduced.

Following work of R.~Jackiw \cite{Jackiw}, we then show in Section 3 that, for $s=2$, and a given value of the ultraviolet cutoff $\kappa$, there are two ways to perform the nrl: i.) fixing the renormalized mass $m^2$ equal to the bare mass and 
ii.) keeping the renormalized mass different from the bare mass, i.e.,  $m^2 = m_0^2$. In the (infinite-volume) 
two-particle sector the scattering amplitude tends to zero as $\kappa \to \infty$ in case i.) and, in case ii.), there is a bound state, indicating that the interaction potential is attractive. 
As a consequence, stability of matter fails for our boson system. 
We discuss why both alternatives do not reproduce the low energy behavior of the full relativistic theory.

R.~Jackiw \cite{Jackiw} treated the problem by the method of the self-adjoint extension, but did not attempt to identify the free parameter occurring there with the physical parameters of the full relativistic theory. We complete 
his work by showing that, in case ii.), the existence of a bound state follows from the two-body interaction 
Hamiltonian obtained in the formal nrl, and that the renormalized mass of the full theory is uniquely related to the free parameter determining the bound state in the nrl.
Case i.) has been considered previously by B\'eg and Furlong \cite{Beg} and K.~Huang \cite{Huang}.
The singular nature of the nrl is also nicely illustrated in Section 4 by a rigorous stability/instability result, 
now for $s=1$, namely, that complete semi-passivity (a notion introduced in \cite{Ku}), which is equivalent to the stability of the vacuum in the presence of an observer in uniform motion to the vacuum's reference system, holds in rqft but is violated by the nrl. Section 5 is devoted to a conclusion and open problems.

\section{The $(: \phi^4:)_{s+1}$ theories and their formal nrl}

\paragraph{The relativistic $: \phi^4:_{s+1}$ model}
Consider the Hamiltonian of a self-interacting scalar field of bare mass $m_0$ in $s$ space dimensions, 
confined to a box of sides $l=(l_1, \ldots, l_s)$ and ultraviolet cutoff $\kappa$ (see \cite{GJ1}\cite{F}) 
(the supscript ${\rm rel.}$ 
stands for `relativistic'):
\beq
\label{1}
H^{\rm rel.}_{l, \kappa} = H^{\rm rel.}_0 + V^{\rm rel.}_{l, \kappa} .
\eeq
As usual, $H^{\rm rel.}_0$ is the kinetic energy operator
\beq
\label{2}
H^{\rm rel.}_0 = \int \d^s \vec p \;  \; a^*(\vec p \, ) \omega_c (\vec p\,) a (\vec p\,) ,
\eeq
where $a(\vec p\,)$, $a^*(\vec p\,) $ are standard annihilation and creation operators  on the symmetric 
Fock space $\cF(\cH)$ over $\cH = L^2 (\rr^s , \d^s \vec x \,)$ satisfying (formally) 
\[    [  a (\vec p \,  ) , a^*(\vec p \, ') ] = \delta (\vec p - \vec p \, ') ,\]
and
\beq
\label{3}
\omega_c (\vec p \, ) = ( \vec p\,^2  c^2 + m_0^2 c^4)^{1/2}.
\eeq
Let $\phi_{c, \kappa}$ denote the (time-zero) scalar field of mass $m_0$ with a sharp ultraviolet cutoff $\kappa$:
\beq
\label{4}
\phi_{c, \kappa} (\vec x) = (2 \pi)^{-s/2} c \int_{| p_i| < \kappa} \frac{\d^s \vec p} { (2 \omega_c (\vec p\,))^{1/2}} \, \e^{-i \vec p \vec x}
\bigr(a^*(\vec p \,) + a (-\vec p \,) \bigl) , \qquad i=1, \ldots, s .\eeq 
The interaction is 
\beq
\label{5}
V^{\rm rel.}_{l, \kappa} = \lambda_0 \int_{|x_i | \le l/2} \d^s \vec x \; 
: \phi^4_{c, \kappa} : \; + \; (\delta m_\kappa)^{2}
\int_{|x_i | \le l/2} \d^s \vec x   \; : \phi^2_{c, \kappa} :, \quad  \; i=1, \ldots, s,
\eeq
(+ eventual renormalization counterterms). 
The term $(\delta m_\kappa)^{2}$ is a mass renormalization counterterm (\cite{GJ1}\cite{F}), 
and the `other renormalization counterterms' are best described, in our context, by Theorem 2.4 and Theorem 2.5
of \cite{SS} and the references given there. These terms will not play any role in our considerations.

Let $E^{\rm rel.}_{l, \kappa}$ denote the infimum of the spectrum of  $H^{\rm rel.}_{l, \kappa}$ on $\cF (\cH)$, 
corresponding to the energy of the `interacting vacuum'. In order that the energy $E^{\rm rel.}_{l, \kappa} >- \infty$ we chose
\beq 
\label{8}
\lambda_0 > 0  
\eeq
in (\ref{5}).
Above, ``$:.:$'' denotes Wick ordering w.r.t.~the Fock
vacuum.

For $s=1$ the ultraviolet limit can be removed without changing the representation \cite{GJ3}: 
The limit
\beq
\label{uvlimit}
\lim_{\kappa \to \infty} H^{\rm rel., s=1}_{l, \kappa}  =: H_l^{{\rm rel.}, s=1}\eeq
exists in the strong resolvent sense, and defines  an essentially self-adjoint unbounded operator $H_l^{{\rm rel.}, s=1}$ on the Fock space $ \cF(\cH)$, which is bounded from below.

\bigskip
We define thermodynamic stability as follows.

\begin{definition} The theory defined by (\ref{1}) is thermodynamically stable if 
$\exists \;  0 < {\tt C}_{\rm rel.} < \infty$ such that 
\beq
\label{7}
E^{\rm rel., s}_{l, \kappa} \ge - {\tt C}_{\rm rel.} \,  l^s, \qquad \kappa  \; \; {\rm fixed} . 
\eeq
Here $E^{{\rm rel.}, s=2}_{l, \kappa}$ (respectively,~$E^{{\rm rel.}, s=1}_{l, \kappa}$) denotes the infimum of the spectrum of 
$H^{{\rm rel.}, s=2}_{l, \kappa}$ (respectively,~$H^{{\rm rel.}, s=1}_{l, \kappa}$) on $\cF(\cH)$, corresponding to the
`interacting vacuum'. 
\end{definition}

The above property is a property of the vacuum state, but it is a necessary condition for the existence of the thermodynamic functions also for nonzero temperature, i.e., in thermal field theory \cite{Jae}, for the same
reason that stability of non-relativistic $n$-particle systems (the forthcoming Definition 2)
is a necessary condition for the existence of the thermodynamic limit (see \cite{Jackiw}, p.~62). 

\paragraph{The formal non-relativistic Hamiltonian}
Take now in (\ref{5})  periodic b.c.~on a segment (s=1) (resp.~a square (s=2)) of length (resp.~side) $l$, denoted
by $\Lambda_l$. By (\ref{2}) and (\ref{5}) taking the thermodynamic limit $l \to \infty$ in the $(\delta m_\kappa)^2$-term in (\ref{5}) 
we obtain an effective kinetic energy
\beq
\label{6neu}
\tilde H_0 := \int \d^s \vec p \; a^*( \vec p\, ) \tilde \omega_c (\vec p\,) a(\vec p\,)
\eeq
with
\beq
\label{7neu}
\tilde \omega_c (\vec p \,) = (\vec p\,^2 c^2 + m^2 c^4)^{1/2} .
\eeq
Here
\beq
\label{8neu}
m^2 := m_0^2 + (\delta m_\kappa)^2 
\eeq
is the renormalized mass. We shall assume that $(\ref{6neu})-(\ref{8neu})$ are valid for $\kappa$ and $l$
sufficiently large, with $m$  (close to) a fixed positive value, the `observed particle mass'. 
Thus, the Hamiltonian (\ref{1}) may be written
\beq
\label{9a-neu}
H^{{\rm rel.}, s}_{l, \kappa} := \tilde H_0 + \tilde V_{l, \kappa},
\eeq
where
\beq
\label{9b-neu}
\tilde V_{l, \kappa} := \lambda_0 \int_{|x_i|\le l/2} \d^s \vec x : \phi^4_{c, \kappa}: + \quad \hbox {\rm (other renormalization counterterms).}
\eeq
By (\ref{7neu}), as $c \to \infty$, we have
\beq
\label{10neu}
\tilde \omega_c (\vec p\,) = mc^2 + \frac{\vec p\,^2}{2m} + O(c^{-2}).
\eeq
Let  
\beq N = \int \d^s \vec p \; a^*(\vec p \,)  a (\vec p\,) \eeq
denote the number operator. We denote by $n=0, 1, 2, \ldots,$ its eigenvalues, whose corresponding eigenspaces are the $n$-particle sectors. We note that for $s=2$ the ultraviolet renormalization involves a change of representation and 
the Fock space number operator is no longer appropriate. We will avoid this problem by keeping the UV cutoff $\kappa$ in case $s=2$.

By (\ref{6neu}), (\ref{9a-neu}) and (\ref{10neu}), there exists in the limit $c \to \infty$ an infinite
`Zitterbewegung' term $N \cdot mc^2$. If we keep $c$ fixed in this term,
the Zitterbewegung term guarantees that the energy-momentum spectrum of the non-relativistic
theory lies inside the relativistic forward light cone, if stability of matter holds with a sufficiently small constant. 
In other words, the nrl satisfies the relativistic spectrum condition.  

Since the Zitterbewegung term (with $c=1$) is constant for fixed particle number, it is usually
 ignored (this formal subtraction being part of the limiting prescription). We thereby obtain from (\ref{4}), (\ref{9a-neu}), (\ref{9b-neu}) and (\ref{10neu}), in the formal limit $c\to \infty$, the non-relativistic many-body Hamiltonian
\beq
\label{12neu}
\begin{array}{rl}
H^{\hbox{\tiny n.-rel., s}}_{l} &= \int \d^s \vec p \; a^*(\vec p \, ) \frac{\vec p^2}{2m} a^*(\vec p \,)+
\\ [3mm]
&
\qquad + \12 \frac{3 \lambda_0}{m_0^2} \int_{| \vec x | \le l/2} \d^s \vec x \int_{| \vec y | \le l/2} \d^s \vec y
\; a^*(\vec x \, )a^*(\vec y \, ) \delta_\kappa^{(s)} (\vec x - \vec y \, ) a(\vec y\, )a(\vec x \, ). 
\end{array}
\eeq
Here $\delta_\kappa^{(s)}$ is the `regularized' delta function whose Fourier transform 
$\tilde \delta_\kappa^{(s)}$ acts on $L^2  (\rr^s, \d^s \vec k \, )$ by
\beq
( \tilde \delta_\kappa^{(s)} \tilde f ) (\vec k\,) =  \int \d^s \vec p \;  \; \chi_\kappa (\vec p \, ) \unit (\vec k- \vec p \,) 
\chi_\kappa (\vec k \,)  \tilde f (\vec p \,), \qquad \tilde f \in L^2  (\rr^s, \d^s \vec k \, ), 
\label{11}
\eeq
with
\beq
\chi_\kappa \ (  \vec q\,) = \chi_\kappa  (  q_1)  \chi_\kappa   (  q_2)  \ldots
\chi_\kappa   (  q_s) 
\label{12}
\eeq
for $\vec q = (q_1, \dots, q_s)$ and 
\beq
\chi_\kappa   (q) =  \Bigl\{ \matrix{ & 1 &  |q| \le \kappa, \cr
& 0 & {\rm otherwise.}
\cr }
\eeq
This is a consequence of  the sharp momentum cutoff in (\ref{5}). 
In (\ref{12neu}), the factor $\frac{3}{2}= \frac{6}{4}$ arises from the number of ways of getting two creators and two annihilators
in the Wick pro\-duct~(\ref{5}).

\paragraph{Remark 2.1}
It is remarkable that Dimock \cite{D} showed convergence
of the two-particle scattering amplitude of the rqft for $s=1$
to the corresponding object for the $\frac{ 3 \lambda_0}{m_0^2} \delta (x)$ potential in
the case $n=2$ (specializing the more general theorem of \cite{D} to our case).
We assume that convergence also holds for general $n$ as conjectured in
\cite{D}(Concluding Remark 2). In correspondence to (2.16) we also assume
convergence of the rqft (2.12) for $s=2$ to the nonrelativistic theory
(2.16) in the same sense described in \cite{D} but only for $\kappa$ smaller than infinity
(the limit $\kappa$ tending to infinity is subtle, see \cite{D'AFT}).
The terms with particle creation and destruction (including the vacuum polarization term) do not automatically drop from (\ref{9a-neu}), (\ref{9b-neu}) in the formal (nrl), but Dimock's work leads us to expect (and assume) that they are absent.  The other renormalization counterterms in (\ref{9b-neu}) yield
(for $0 < \kappa < \infty$) constants, as $c \to \infty$, which do not affect the discussion of  thermodynamic stability
(see~\cite{SS}, Theorem~2.4 and 2.5, and Theorem 6.2). We shall thus ignore them henceforth. 

\bigskip
The formal limit (\ref{12neu}) commutes with the number operator $N$
and thus leaves each $n$-particle sector (where $n=0,1, \ldots, $   is an eigenvalue of $N$)  invariant. 
The Hamiltonian (\ref{12neu}) corresponds to the two-body Hamiltonian
\beq
\label{f2.20a}
H_{\kappa, 2}^{\hbox{\tiny n.-rel., s} } = - \frac{1}{2m} \Delta + V_\kappa^{(s)} \; , 
\eeq
where $\Delta$ is the Laplacian, 
\beq
\label{f2.20b}
V_\kappa^{(s)}:= \frac{3 \lambda_0}{ m_0^2} \delta^{(s)}_\kappa (\vec x \, ) \, ,
\eeq
and $\delta^{(s)}_\kappa$ is given by (\ref{11}).

We shall consider (as in \cite{D}, for $s=1$ and $\kappa = \infty$) the above Hamiltonian in infinite space, i.e., as
an operator on $L^2 (\rr^s)$. The corresponding Hamiltonian restricted to the $n$-particle sector is 
\beq
\label{f2.21}
H_{\kappa, n}^{\hbox{\tiny n.-rel., s} } = - \frac{1}{2m} \sum_{i=1}^n \Delta_i + \sum_{1 \le i < j \le n } V_\kappa^{(s)}
(\vec x_i - \vec x_j  \, )
\eeq
on $L^2(\rr^{sn})$. 

Next we consider a notion of stability which usually is referred to as ``stability of matter''.

\begin{definition}Let $E^{\hbox{\tiny n.-rel., s} }_{\kappa,n}$ denote 
the infimum of $H_{\kappa,n}^{\hbox{\tiny n.-rel.,s} }$ in the $n$-particle sector.
The non-relativistic Hamiltonian $H^{\hbox{\tiny n.-rel.,s}}_{\kappa,n}$ is stable if $\exists \; 0 < 
{\tt C}_{\hbox{\tiny n.-rel., }\kappa } < \infty$ such that 
\beq
\label{14}
E^{\hbox{\tiny n.-rel., s} }_{n,\kappa} 
\ge - {\tt C}_{\hbox{\tiny n.-rel.,} \kappa }  \; n \quad \forall n \in \nn . 
\eeq
\end{definition}

It is important to remark that, formally, the density 
$\rho = \frac{n}{l^s}$ corresponding to the nonrelativistic Hamiltonian $H_{\kappa, n}^{\hbox{\tiny n.-rel., s} } $
is zero (in Ref.~\cite{D}, for instance, $s= 1$; $n=2$ and $l= \infty$). For $s=1$ and $\kappa= \infty$ we have
that (\ref{f2.21}) becomes
\beq
\label{f2.23}
H_{\kappa, n}^{\hbox{\tiny n.-rel., s=1} } = - \frac{1}{2m} \sum_{i=1}^n \frac{\partial^2}{\partial x_i^2}
+ \frac{3 \lambda_0}{m_0^2} \sum_{1 \le i < j \le n } \delta 
( x_i -  x_j    ) .
\eeq
The above formal expression (\ref{f2.23}) may be made precise in a standard way (Example X.3 of \cite{RS}). 
Taking, however, (\ref{f2.23}) with periodic b.c.~on a segment of length $l$
the corresponding quantity $H_{n, l}$ too may be defined for arbitrary $n \in \nn$
as in \cite{Do} and in this case one branch of 
the correponding elementary excitations of momentum $p$ is, in the hard core case, 
(see \cite{LiLi} Part II, Equ.~(2.8), p.~1618)
\[ \epsilon_1 (p) = p^2 + 2 \pi \rho | p|  \qquad  (2m=1);  \]
this agrees with the free case (only) if $\rho =0$.

All this makes it clear that there is a certain weak point in our comparison: the relativistic neutral scalar 
$:\phi^4:$ models do not allow us to vary the particle density independently from the temperature. Thus it 
would be interesting to investigate the non-relativistic limit of 
charged sectors in the $: \phi \overline {\phi} :^2$ model, where the charge (density) can be kept constant.

The definition of the regularized delta  function (\ref{11}) does not make the (in)stability properties of the corresponding two-body potential explicit: 
for that purpose, it is judicious to search a representation in configuration space. We present two versions of the 
two-body potential in infinite volume. 

\paragraph{The lattice approximation in infinite volume}
Let 
$B_{\epsilon} := [ - \frac{\pi}{{\epsilon}},  \frac{\pi}{{\epsilon}}]^2$ denote the first Brillouin zone of an (infinite) lattice of spacing 
${\epsilon}$ along the coordinate axes. The (lattice-cutoff) Hilbert space is $\cH_{\epsilon} := l^2 (\zz^s)$. 
On $\cH_{\epsilon}$ we define the Hamiltonian corresponding to~(\ref{12neu}) in the 2-particle sector. 
Separating the center of mass Hamiltonian, the Hamiltonian in  relative coordinates is
\beq
\label{19a}
H_{\epsilon} := H_{0, {\epsilon}} + V_{\epsilon}  \, ,
\eeq
where
\beq
\label{19b}
(H_{0, {\epsilon}} \Psi ) (\vec k) = E_{\epsilon} (\vec k) \Psi (\vec k)  
\eeq
with
\beq
\label{19c}
E_{\epsilon} (\vec k) = {\epsilon}^{-2} \sum_{i=1}^{2} ( 1 - \cos {\epsilon} k_i), \qquad \vec k = (k_1, k_2),
\eeq
and
\beq
\label{19d}
(V_{{\epsilon}} \Psi ) (\vec k) = \frac{3 \lambda_0}{m_0^2 }  \int_{B_{\epsilon}} \d^2 \vec k \; \; \Psi (\vec k) .
\eeq
Note that (\ref{19d}) corresponds precisely to (\ref{12neu}) and (\ref{11}) on the lattice, with 
\beq
\label{star}
\kappa 
= \pi / \epsilon.
\eeq

\paragraph{The square-well regularization} 
We replace $V_{\epsilon}$ in (\ref{19d}) by the potential 
\beq
V^{\epsilon} \bigl(  \vec x  \bigr) := \frac{3 \lambda_0}{m_0^2 }  \Phi^{\epsilon} \bigl(  \vec x  \bigr)
\label{20a}
\eeq
where 
\beq
\Phi^{\epsilon} \bigl(  \vec x  \bigr) := \chi_{\epsilon} \bigl( | \vec x_i - \vec x_j | \bigr) \frac{2}{{\epsilon}^2} 
\label{20b}
\eeq
and
\beq
\chi_{\epsilon} \bigl( | \vec x_i - \vec x_j | \bigr) = \Bigl\{ \matrix{ & 1 & {\rm if} \quad 0 \le | \vec x | \le {\epsilon},  \cr
&  0 & {\rm otherwise}.&    \cr} 
\label{20c}
\eeq
We shall identify $\kappa$ in (\ref{f2.21}) with (\ref{star}) also for (\ref{20a}). This is true
(up to a multiplicative factor), see ((3.35),(3.40) and the remark
thereafter in Ref. \cite{Jackiw}). This assertion means that bound state energies and
scattering amplitudes agree in both models (see also Ref. \cite{DY}).
In the following, we shall thus feel free to use the square-well regularization as a prototype to prove our main results.
\bigskip

\section{The formal nrl for $s=2$} 

Consider the two-body Hamiltonian (\ref{f2.20a}). If $0 < \kappa < \infty$, the nrl may be performed in two ways: 
i.)~by insisting that the mass renormalization constant $(\delta m_\kappa)^2 = 0$ in (\ref{8neu}), i.e., $m^2 = m_0^2$; ii.)~by fixing the physical mass $m^2$ in 
(\ref{8neu}), i.e., $(\delta m_\kappa)^2 \to \infty$ for $\kappa \to \infty$.
Accordingly we have the following result.

\begin{proposition} In case i.) the scattering amplitude for (\ref{f2.20a}) tends to zero as $\kappa \to \infty$. 
In case ii.) there exists  $\kappa_0$ such that, if $\kappa_0 < \kappa < \infty$, there exists a bound state, the coupling 
constant is negative and (thus)
the corresponding bosonic Hamiltonian (\ref{f2.21}) is unstable according to Definition 2.2.
\end{proposition}

\noindent
\proof 
In case i.), by (\ref{8neu}) and (\ref{f2.20b}), the scattering solutions $\tilde f$ corresponding to the potential in 
(\ref{11}) in the two-particle sector satisfy the Lippman-Schwinger equation (we choose an informal presentation following closely 
\cite{Jackiw} in order to avoid unnecessary technicalities, but the results agree completely with those in Chapter I.5
of \cite{AGHKH}, obtained by the method of self-adjoint extensions). We start from the Schr\"odinger equation:
\beq
\label{f3.1} \frac{1}{2m_0} ( \vec p \,^2 - \vec k^2) \tilde f (\vec p \, ) = -  \frac{3 \lambda_0}{m_0^2}  {\bf V} ( \tilde f, \kappa)
\eeq
where 
\beq
\label{f3.2}  {\bf V} ( \tilde f, \kappa) := \int_{| \vec p_i | \le \kappa; \, i=1,2} \frac{\d^2 \vec p}{(2 \pi)^2} \tilde f (\vec p \, ) 
\eeq
and 
\beq
\label{f3.3} \frac{\vec k^2 }{2 m_0} = E > 0 
\eeq
is the energy of the scattered particles. From (\ref{f3.1}) we get the Lippman-Schwinger equation
\beq
\label{f3.5} \tilde f (\vec p \, ) = (2 \pi)^2 \delta (\vec p - \vec k) - \frac{2 }{\vec p\,^2 - \vec k^2 -i \epsilon}
\frac{3 \lambda_0}{m_0^2} {\bf V} ( \tilde f, \kappa)  \, \, .
\eeq
By (\ref{f3.5}) the scattering amplitude is proportional to $\frac{3 \lambda_0}{m_0^2}  {\bf V} ( \tilde f, \kappa)$. Integrating (\ref{f3.5}) over the region $| \vec p_i | \le \kappa$, $ i=1,2$, we get for $| \vec k_i | \le \kappa$, $ i=1,2$
\beq
\label{f3.6a}   \int_{| \vec p_i | \le \kappa; i=1,2} \frac{\d^2 \vec p}{(2 \pi)^2} \tilde f (\vec p \, ) = {\bf V} ( \tilde f, \kappa) 
= 1
- 2 {\bf I} (- \vec k^2 - i \epsilon)  \frac{3 \lambda_0}{m_0^2} {\bf V} ( \tilde f, \kappa) 
\eeq
with 
\beq
\label{f3.7} {\bf I} (z)  = \int_{| \vec p_i | \le \kappa; \, i=1,2} \frac{\d^2 \vec p}{(2 \pi)^2} \frac{1}{\vec p \,^2 + z}.
\eeq
If some $| \vec k_i | > \kappa$, $ i=1,2$ the $1$ on the r.h.s.\ is absent in (\ref{f3.6a}). 
Equivalently
\beq
\label{f3.6b} \frac{3 \lambda_0}{m_0^2}  {\bf V} ( \tilde f, \kappa) = \Bigl( \frac{m_0^2}{ 3 \lambda_0} + 2 {\bf I}  (- \vec k^2 - i \epsilon) \Bigr)^{-1}, \qquad \hbox{for $| \vec k_i | \le \kappa$, $ i=1,2$}.
\eeq
We have, asymptotically for large $\kappa$ (using condition $| \vec p \, | \le \kappa$ in (\ref{f3.7}) instead 
of $| \vec p \, | \le \kappa$, $i=1,2$):
\beq
\label{f3.8} {\bf I} (z)  =  \frac{1}{4 \pi} \ln \frac{\kappa}{ z} \, , 
\eeq
and thus, from (\ref{f3.6b}), 
\beq
\label{f3.9} \frac{3 \lambda_0}{m_0^2}  {\bf V} ( \tilde f, \kappa) = \left(
\frac{m_0^2}{ 3 \lambda_0}
+ \frac{1}{\pi} \ln \frac{\kappa}{\mu} - \frac{1}{ \pi} \ln \frac{ | \vec k |}{\mu} + \frac{i \epsilon}{2} \right)^{-1}
\eeq
where $\mu$ is a renormalization point. 
Since, from (\ref{f3.5}), $\frac{3 \lambda_0}{m_0^2} {\bf V} ( \tilde f, \kappa)$ is the scattering amplitude, we have from (\ref{f3.9}) that it tends to zero as 
$\kappa \to \infty$, which is the first result. This observation is already contained in \cite{Jackiw}, 
\cite{Beg} and \cite{Huang}.

Concerning ii.) we also follow \cite{Jackiw} but, instead of appealing to the method of self-adjoint extension with a free parameter we use (\ref{f2.20a}), replacing, however, $m_0^2$ in the denominator of (\ref{f2.20b}) by equation
(\ref{8neu})
\beq
m_0^2 = m^2 - \delta m_\kappa^2
\label{f3.10}
\eeq
with $m^2$ fixed. By \cite{GJ1}\cite{F}
\beq
\delta m_\kappa^2 = O \left( \ln \frac{1}{\kappa} \right) 
\label{f3.11}
\eeq
(i.e., $\lim_{\kappa \to \infty} \frac{\delta m_\kappa^2 }{ \ln \kappa} = c >0 $). By (\ref{f3.10}) and (\ref{f3.11}), 
with (\ref{star})--(\ref{20c}), the Hamiltonian (\ref{f2.20a}) tends, in the norm resolvent sense, to a self-adjoint point interaction Hamiltonian as $\epsilon \to 0$ (or $\kappa \to \infty$), the ``$ \delta^{(2)}(\vec x \, )$'' with the precise 
properties which we shall now use, following \cite{Jackiw} (\cite{AGHKH}, Chapter I.5, (5.47) and Theorem 5.5).
These properties are equivalent to the following ``field theoretic'' language \cite{Jackiw}: taking the ``bare coupling''
(here of the formal nrl, i.e., the coupling $\frac{3 \lambda_0}{m_0^2}\;$) to be cutoff-dependent, we may now introduce a 
``renormalized coupling contant $g$'' (again of the nonrelativistic theory) in terms of which (\ref{f3.9}) may be written
\beq
\label{f3.10a}
\frac{3 \lambda_0}{m_0^2}  {\bf V} ( \tilde f, \kappa)= \Bigl( \frac{1}{g} - \frac{1}{\pi} \ln \frac{| \vec k |}{\mu} + \frac{i \epsilon}{2} \Bigr)^{-1}
\eeq
which is finite and well-defined, provided the factor $1/\frac{3 \lambda_0}{m_0^2}$ absorbs by definition the cutoff dependence~\cite{Jackiw}:
\beq
\label{f3.11a}
\frac{1}{g}  = \frac{m_0^2}{ 3 \lambda_0} + \frac{1}{\pi} \ln \frac{\kappa}{\mu}  .
\eeq
From here we proceed as in \cite{Jackiw}, but for the benefit of the reader, we summarize the main steps.
Write the Lippman-Schwinger equation (\ref{f3.5}) in configuration space (see also \cite{AGHKH}, Chapter I.5):
\beq
\label{f3.12}
f(\vec x \, )  = \e^{i \vec k \vec x } -  \frac{6 \lambda_0}{m_0^2} G_k \bigl( | \vec x | \bigr)  {\bf V} ( \tilde f, \kappa)
\eeq
where $G_k \bigl( | \vec x | \bigr)$ is the Green's function appropriate to two dimensions,
\beq
\label{f3.13}
(- \Delta - \vec k^2)   G_k \bigl( | \vec x | \bigr) = \delta^{(2)} (\vec x \, ) \, , 
\eeq
and, thus,
\beq
\label{f3.14}
G_k \bigl( | \vec x | \bigr) 
\to
 \frac{  \e^{i  ( | \vec k | \, | \vec x | + \pi /4 )} } { 2 \sqrt{ 2 \pi | \vec k | \, |  \vec x |}}
\quad \hbox{for \ }   | \vec x| \to \infty . 
\eeq
From 
\[ f (\vec x \, ) \sim \e^{i \vec k \vec x } + \frac{1}{\sqrt{ | \vec x|}} {\bf f} (\theta) \e^{i ( | \vec k | |\vec x | + \pi /4 ) } \, , \]
(\ref{f3.12}),  and (\ref{f3.14}) we obtain
\beq
\label{f3.15}
{\bf f} (\theta) = - \frac{1}{\sqrt{2 \pi |\vec k |}} \frac{3 \lambda_0}{m_0^2}  {\bf V} ( \tilde f, \kappa) =  - \frac{1}{\sqrt{2 \pi |\vec k |}} \, 
\left(\frac{1}{g} - \frac{1}{\pi} \ln \frac{ |\vec k | } {\mu} + \frac{i \epsilon}{2} \right)
\eeq
and thus only s-wave scattering takes place with phase shift $\delta_0$ such that $\cot \delta_0 = \frac{1}{\pi}
\ln \frac{|\vec k |^2}{ \mu} - \frac{2}{g}$; it may also be seen \cite{Jackiw} alternatively to $g$, that one may fix a bound state energy $B$ ($E=-B$) such that
\beq
\label{f3.16}
{\bf f} (\theta) = - \frac{1}{\sqrt{2 \pi |\vec k |}}   
\Bigl( \frac{1}{2\pi} \ln \frac{ 2B} {|\vec k |^2} + \frac{i \epsilon}{2} \Bigr)^{-1}
\eeq
with 
\beq
\label{f3.17}
\sqrt{2B} = \mu \, \e^{\pi / g} .
\eeq
By (\ref{f2.20b}), (\ref{f3.10}) and (\ref{f3.11}) we see that the point-interaction in two space dimensions is always attractive
(for fixed $0 < m^2 < \infty$ and $\kappa$ sufficiently large), and thus,  since there are no additional repulsive interactions 
(no ``hard core''), the hamiltonian is unstable according to the Definition 2.2 by standard arguments (\cite{D'AFT}, p.62).
We expect that the Hamiltonian converges in norm-resolvent sense to the n-body interaction Hamiltonian defined in 
\cite{D'AFT}, which is also unstable according to Definition 2.2. Finally, (\ref{8neu}) and (\ref{f3.11}), 
$m^2$ may be identified with $( \frac{1}{g} ) $, and is, thus, not surprisingly, the free parameter of the full
(relativistic) theory which occurs in the method of self-adjoint extension (see \cite{AGHKH}\cite{Jackiw}). Thus  we see 
that $g$ must be chosen positive, which is not an a priori consequence of (\ref{f3.17}). This completes 
the proof of Proposition 3.1. 

\paragraph{Remark 3.1}
Alternative i.)~corresponds to setting $\delta m_\kappa^2 \equiv 0$ by (\ref{8neu}). It may be argued that the mass
counterterm is created solely to cancel diagrams in which the particle number 
changes (see \cite{GJ1}) and since the latter terms are expected to vanish in the nrl, it would be consistent to require assumption
i.). On the other hand, Proposition 3.1 shows  that one thereby obtains zero scattering amplitude in the formal nrl, which does not correspond to the low-energy behavior of the full theory, which is nontrivial  (\cite{GJ1}\cite{F}).

Alternatively ii.)~corresponds to the principle that although the function of the mass term is to cancel particle-changing
diagrams which do not appear in the nrl, it also fixes the physical parameters of the full 
$(: \phi^4:)_3$ theory (i.e., with all the terms, including those changing particle numbers), and thus must be kept 
fixed upon performing the formal nrl. We see this clearly from the fact that the $\delta m_\kappa^2$-term contributes
to the free (kinetic) energy, yielding the renormalized mass $m$. In fact, it is not $\sqrt{p^2 c^2 + m_0^2 c^4}$
that is expanded to obtain the nonrelativistic 
kinetic energy, but $\sqrt{\vec p\,^2 c^2 + m^2 c^4}= mc^2 + \frac{\vec p\,^2}{2m} + O(c^{-2})$, see~(\ref{10neu}).
In other words, it is particles with mass $m$ that scatter, also in the nonrelativistic limit! This is also the basis of Bethe's nonrelativistic treatment of the Lamb shift \cite{Sakurai}. Finally, the $\delta m_\kappa^2$-term in (\ref{5}) also contains particle conserving terms. 

Although the above arguments favor, in our opinion, assumption ii.), the latter also yields a result in complete disagreement with the low-energy behavior of the full theory. Indeed the rqft is stable according to Definition 2.1
(a consequence of Nelson's symmetry, see, e.g.,  \cite{SS}). The interaction is expected to be of purely repulsive character, and in particular no bound states are expected to exist in the full rqft \cite{4point}. Although the latter reference pertains to the case $ s=1 $, the methods used ,in particular, Lebowitz's inequality, should be applicable to show the same assertion for $ s=2 $. 
 We comment further on the above mentioned disagreement in the last section.
 
Eliminating the ``Zitterbewegung'' term already accounts for the singular nature of the nrl, which is responsible for instability results of a different nature, which we now examine.

\section{Passivity and Local Thermodynamic Stability}

Let us consider a $C^*$-dynamical system, consisting of a $C^*$-algebra $\cA$ and a one-parameter group of automorphisms $\{ \tau_t \}_{t \in \rr}$. When the time-evolution
$t \mapsto \tau_t \in Aut (\cA)$ is changed by a local perturbation, which is slowly switched on and
slowly switched off again, then an equilibrium state 
returns to its original form at the end of this procedure. This 
heuristic condition of {\sl adiabatic invariance} can be  expressed by the
stability requirement 
\beq \lim_{t \to \infty} \int_{-t}^t \d t \thinspace \omega \bigl( [ a, \tau_t(b)] \bigr) = 0
\qquad \forall a, b \in  \cA. \label{adiabatic}\eeq
In a pioniering work Haag, Kastler and Trych-Pohlmeyer \cite{HKT-P} showed that
the characterization (\ref{adiabatic}) of an equilibrium state 
leads to a sharp mathematical criterion,
first encountered by Haag, Hugenholtz and Winnink \cite{HHW} and more 
implicitly by Kubo \cite{K}, Martin and Schwinger \cite{MS}:

\begin{definition}{A state
$\omega_\beta$ over a $C^*$-algebra $\cA$ is called a KMS state for 
some $\beta >0$, if for all $a, b \in \cA$, there exists a function  $F_{a,b}$ which is continuous in
the strip $0 \le \Im z \le \beta$ and analytic and bounded  in the open strip
$0 < \Im z < \beta$, with boundary values given by
\[ F_{a,b}(t) = \omega_\beta \bigl( a \tau_t (b) \bigl) 
\quad 
\hbox{and}
\quad
F_{a,b}(t + i\beta) = \omega_\beta \bigl( \tau_t (b) a \bigl)  \quad \forall t \in \rr.\]
}\end{definition}

The amount of work a cycle can perform when applied to a moving thermodynamic equilibrium state is bounded by the amount of work an ideal windmill or turbine could perform; this property is called {\em semi-passivity} \cite{Ku}:
A state $\omega$ is called semi-passive  (passive) if there is an `efficiency bound' $E \ge 0$ ($E=0$)
such that 
\[ - (W \Omega_\omega, H_\omega W \Omega_\omega)  \le E  \cdot ( W \Omega_\omega , |P_\omega| W \Omega_\omega)  
 \]
for all unitary elements $W \in \pi_\omega (\cA)''$, which satisfy
\[ [H_\omega, W] \in \pi_\omega (\cA)'', \qquad [P_\omega, W] \in \pi_\omega (\cA)'' . \]
Here $(H_\omega, P_\omega)$ denote
the generators implementing the space-time translations in the GNS representation $(\cH_\omega, \Omega_\omega, \pi_\omega)$.  

Generalizing the notion of complete passivity (which is related to the zeroth law of thermodynamics), the state $\omega$
is called {\em completely semi-passive} if all its finite tensorial powers are semipassive 
with respect to one fixed efficiency bound $E$. It has been shown by Kuckert \cite{Ku} that
a state is completely semi-passive in all inertial frames if and only if it is completely passive in some inertial frame.
The latter implies that~$\omega$ is a KMS-state or a ground state (a result due to Pusz and Woronowicz). 
We shall be interested in the latter case. 

For ground states (characterized by the positivity of the energy operator $H \ge 0$) the 
stability property (\ref{adiabatic}) implies the existence of a mass gap:

\begin{proposition} {\rm (Bratteli, Kishimoto, Robinson \cite[Theorem 3]{BKR})}. Let $(\cA, \tau)$ be a
$C^*$-dynamical system and let $\omega$ be a strongly clustering $\tau$-ground state, i.e.,
\beq \lim_{t \to \infty} \omega \bigl(  a \tau_t(b)  \bigr) = \omega (a)\omega (b) 
\qquad \forall a, b \in  \cA. \eeq
Let $U_\omega (t) = \e^{it H_\omega}$ be the unitary group which implements $\tau$ in the 
GNS representation $(\cH_\omega, \pi_\omega, \Omega_\omega)$. 
Then the condition (\ref{adiabatic}) is 
equivalent to the existence of some $\epsilon>0$ such that
\[ {\rm sp} \; H_\omega \subset \{ 0 \} \cup [ \epsilon, \infty [ \; ,\]
where ${\rm sp} \; H_\omega$ denotes the spectrum of $H_\omega$.
\end{proposition}

By \cite[Theorem 2]{WFW} or \cite[Theorem 1]{RWFW}, if the dynamics~$\tau_t$ is
nontrivial, i.e., $H_\omega $ is not $0$, the extension of $\omega$ to $\pi_\omega (\cA)''$ is
nonfaithful and completely passive if and only if it is a ground state
(see Footnote 1).  The following proposition of \cite{Ku} is thus applicable:

\begin{proposition}{\rm (\cite[Proposition 3.2]{Ku}).}
Let $\omega$ be completely semi-passive with efficiency bound $E$. If $\omega$ is not faithful, then there exists
some $\vec u \in \rr^s$, $|\vec u |\le E$, such that 
\beq H_\omega + \vec u \vec P_\omega \ge 0.
\label{staku}
\eeq
\end{proposition}

Proposition 4.3 only asserts the existence of some $\vec u$, with $|\vec u| \le E$, satisfying (\ref{staku}), 
but it may well be that only $\vec u =0$ satisfies (\ref{staku}).
This is however not the case in rqft. Consider the $(: \phi^4:)_2$ theory; we omit the superscript $s=1$ henceforth. 

\begin{proposition} The inequality (\ref{staku}) holds for the $(: \phi^4:)_2$ theory for all $| \vec u | < c$.
\end{proposition}
\proof
According to Heifets and Osipov \cite{HO}
\[ H_l^r -E_l^r \pm v P_l \ge 0 , \qquad 0 \le |v| < c, \]
where $P_l$ is the momentum operator with periodic boundary conditions on a segment of length $l$. This yields
(\ref{staku}). Note that the ground state energy $E_l^r$ is absorbed in  $ H_\omega$.

In fact, since $H_\omega + \vec u \vec P_\omega$ is the Lorentz rotated Hamiltonian, (\ref{staku})
may be expected in general for rqft. Does it hold for its formal nrl? The answer is negative: 

\begin{proposition}
The inequality (\ref{staku}) does not hold for the formal nrl of $(: \phi^4:)_2$ unless~$\vec u= 0$.
\end{proposition}
\proof
The proof uses an idea of G.L.~Sewell. $H_l^{\hbox{\tiny n.-rel.} }$ is Galilean covariant: let
\beq U_{\Lambda_l}^{\pm u} = \e^{-i [t P_l \mp u (x_1 + ...x_n) ] }\eeq
denote the operator implementing Galilean transformations on $L^2 (\Lambda_l)$ (with $u= \frac{2 \pi k }{l} $, $k \in \zz$), 
then
\beq
\label{Se1}
(U_{\Lambda_l}^{\pm u} )^* H_l^{\hbox{\tiny n.-rel.} } U_{\Lambda_l}^{\pm u} = H_l^{\hbox{\tiny n.-rel.} } \pm u P_l + \frac{1}{2} n u^2;
\eeq
\beq
\label{Se2}
(U_{\Lambda_l}^{\pm u} )^* P_l \, U_{\Lambda_l}^{\pm u} = P_l \pm n u .
\eeq
Now assume that
\beq
\label{Se3}
H_l^{\hbox{\tiny n.-rel.} } - E_l^{\hbox{\tiny n.-rel.} } + v P_l \ge 0 \eeq
for some $v>0$. Let $\Omega_{\Lambda_l}$ be the (unique) ground state of 
$H_l^{\hbox{\tiny n.-rel.} }$ corresponding to the eigenvalue $E_l^{\hbox{\tiny n.-rel.} }$, then
\beq
\label{Se4}
P_l \, \Omega_{\Lambda_l} = 0 .
\eeq
Using now (\ref{Se1}), (\ref{Se2}) and (\ref{Se4}) with $u=-v$ we obtain
\beq
\label{Se5}
(U_{\Lambda_l}^{-v} )^* (H_l^{\hbox{\tiny n.-rel.} } - E_l^{\hbox{\tiny n.-rel.} } + v P_l) U_{\Lambda_l}^{-v} \Omega_{\Lambda_l}
= - \frac{1}{2} n v^2 \Omega_{\Lambda_l},
\eeq
which contradicts (\ref{Se3}). Note that the term on the r.h.s.~of (\ref{Se5}) depends on the group parameter~$v$, 
and different $v$'s would change the correction. 

\paragraph{Remark 4.1} The fact that (\ref{staku}) is broken in the nrl is a consequence of the 
singular character of the limit $c \to \infty$, which is a group contraction
(from the Poincar\'e to the Galilei group, see E.~In\"on\"u and E.P.~Wigner \cite{IW}
and U. Cattaneo and the first named author \cite{CWFW}). 
In fact, under the limit $c \to \infty$, the closed foward light cone $\{ (t, \vec x\, ) \in \rr^{s+1} \mid 0 \le t \le  |\vec x \,| /c  \}$ tends to the positive time-axis $\{ (t, \vec 0) \in \rr^{s+1} \mid t\ge 0 \} $. 
Thus it is conceivable that only the vector
$\vec u = \vec 0$ remains in (\ref{Se3}), and that is what Proposition~4.5 demonstrates. This hinges
essentially on 
ignoring the Zitterbewegungsterm. The latter is also responsible for the fact that the mass gap, proved in the full theory for $s=1$ and small coupling in \cite[Theorem 2.2.3]{GJSa}, is made to vanish (at zero momentum) when performing 
the nrl. This has the consequence that the stable vacuum leads to an unstable ground state in the nrl, according to Proposition 4.2.

\section{Concluding Remarks and Open Problems}

A last important structural difference between vacua and ground states concerns the Reeh-Schlieder theorem.
It has been proved by Requardt \cite{Re} that for a class of $n$-body non-relativistic systems, states localized at time zero 
in an arbitrary small open set of $\rr^n$ are already total after an arbitrary small time - a statement which is much stronger
than the well known acausal behavior of non-relativistic theories. However, in contrast to rqft, in non-relativistic theories 
annihilators of the vacuum are affiliated to the von Neumann algebra associated to  bounded regions in coordinate space.

While the (smeared out) field operator of a relativistic theory is affiliated to the von Neumann algebra associated to a bounded space-time region, the (smeared out) creation and annihilation operators 
are not well-localized. Formally, the (time-zero) scalar field of mass~$m_0$ is given by
\beq
\phi_{c} (\vec x) = (2 \pi)^{-s/2}  \int {\d^s \vec p} \; \; \e^{-i \vec p \vec x}
\left(a^* \left(\frac{c \vec p}{ (2 \omega_c (\vec p\,))^{1/2}}  \right) + a 
\left(\frac{ - c \vec p}{ (2 \omega_c (\vec p\,))^{1/2}} \right)\right) .
\eeq 
It is remarkable that the factor 
\[ \frac{c}{ (2 \omega_c (\vec p\,))^{1/2}} \] 
converges to a constant as $c$ goes to
infinity. Consequently, the relativistic annihilation operators, which can not be localized in bounded 
space-time regions, are strictly localized in a bounded space region at sharp time in the limit $c \to \infty$. This finding is
in agreement with the approximation of the non-local creation and annihilation operators by  strictly localized
operators in \cite{GJ4}. It is remarkable that the results in \cite{GJ4} provide  bounds (in an appropriate norm), which 
allow us to control this convergence quantitatively.

It is interesting to note that the proof of stability of the interacting $(:\phi^4:)_3$ theory according  to Definition 2.1 in, 
e.g., Ref. \cite{SS}, makes use of Nelson's symmetry (see \cite{S}, \cite{SS} and references given there), which is itself a consequence of the Lorentz covariance of the full rqft. The latter depends crucially of course on keeping all
particle-changing terms,  including those contributing to vacuum polarization, in the interaction. 
On the other hand, the result of Proposition 3 leads us to conclude that the formal nrl may lead either to
a trivial theory or to an unstable one. This suggests that the elimination of particle-changing terms occurring while performing the nrl may introduce a nonphysical instability which is not present in the full rqft. Since, for $s=2$, there is no Thomas effect \cite{D'AFT} (a special effect related to point interactions), one might conjecture that this phenomenon could 
also take place in other systems of greater physical relevance. 

It should be, however, emphasized that a theory which fails to be thermodynamically stable should not be disregarded 
immediately. It merely shows that, given a set of initial conditions there is a tendency of the system to move out of the domain of validity of the theory. This is a phenomenon well-known from classical general relativity, where the famous singularity theorems of Hawking and Penrose show that for a large class of initial conditions,  the physical system develops,
as a consequence  of the basic thermodynamic instability, hot spots with energy densities which can no longer be described by classical Einstein gravity. In fact, the ``$\delta^{(2)}( \vec x \, )$''- interaction has arisen in discussions of point-particle dynamics in (2+1)-dimensional gravity \cite{HH}.

The phenomenon of instability (nrl) / stability (rqft) is independent of neglecting the ``Zitterbewegungsterm'', 
because it would not be altered if we used  instead the relativistic kinetic energy. Neglecting this term is, however, responsible for the singular nature of the nrl, and accounts for several instability results of a different nature for the resulting
ground states, as shown in Section 4. 

Since the mass or number operator is not conserved in rqft, the density cannot be fixed performing the nrl.
A major open problem is, thus, to consider the charged $(: \phi \overline{\phi} :)^2$-theory, and perform the 
nrl in a sector of fixed charge density. 

\paragraph{Acknowledgement} W.F.W. would like to thank K.~Fredenhagen and B.~Kuckert
for illuminating discussions. Ch.J. is grateful to W.~Wreszinski 
and the Dept.~de Fisica Matematica (USP) for the kind hospitality provided during a visiting 
professorship at the University of S\~{a}o Paulo, supported by FAPESP.

\bigskip

$^{1}${\footnotesize The stronger Theorem 3 of [23] uses norm continuity of $\tau_t$ and is applicable to quantum lattice systems and nonrelativistic and relativistic fermions, but not to the (nonrelativistic and relativistic) bosons treated in the present paper, because of the well-known fact that time-translation automorphisms are not norm continuous on the Weyl algebra. However, the $W^*$-version is elementary: $\forall t \in \rr$, $U_t = \e^{it H_\omega} \in \pi_\omega (\cA)''$
(see \cite[Theorem 3.5 for a simple proof]{HH}). Since $H_\omega \ne 0$, $(U_t - \unit) = \e^{it H_\omega} - \unit \ne 0$
if $t \ne 0$, and $(U_t - \unit) \Omega_\omega = 0$.}

\end{document}